%% file: STDM-2015-22.tex
\pdfoutput=1
\newcommand*{\ATLASLATEXPATH}{./}
\documentclass[UKenglish,texlive=2011,cernpreprint]{\ATLASLATEXPATH atlasdoc}

\usepackage[backend=biber,minimal]{\ATLASLATEXPATH atlaspackage}
\usepackage{\ATLASLATEXPATH atlasbiblatex}
\usepackage{\ATLASLATEXPATH atlasphysics}
\usepackage{wasysym} 
\usepackage{siunitx} 

\graphicspath{{./}}

\usepackage{STDM-2015-22-defs}

\input{STDM-2015-22-metadata}

\hypersetup{pdftitle={ATLAS document},pdfauthor={The ATLAS Collaboration}}

\begin{document}

\maketitle
\section{Introduction}
\label{sec:intro}
The total cross section for proton--proton ($pp$) interactions characterizes a fundamental process of the strong interaction. 
Its energy evolution has been studied at each new range of centre-of-mass  energies available. 
ATLAS has previously reported a measurement of the total cross section in $pp$ collisions at 
$\rts=7 \tev$ \cite{alfa_pub_7}. 
This paper details a measurement of the total cross section at $\rts=8 \tev$ using data collected in 2012. 
The measurement methodology and analysis 
technique are very similar between the two measurements and the technical details are discussed 
thoroughly in Ref. \cite{alfa_pub_7}.  

Both measurements rely on the optical theorem:
\begin{linenomath*}
\begin{equation}\label{eq:OpticalTheorem}
\sigmatot = 4\,\pi \mbox{Im} \, [ f_{\mathrm{el}}\left(t \rightarrow 0\right)] \, 
\end{equation}
which relates the total $pp$ cross section $\sigmatot$ to the elastic-scattering amplitude extrapolated 
to the forward direction $f_{\mathrm{el}}(t \rightarrow 0)$,  
with $t$ being the four-momentum transfer squared. The total cross section can be extracted in different ways using 
the optical theorem. 
ATLAS uses the \textit{luminosity-dependent} method which requires a measurement of 
the luminosity in order to normalize the elastic cross section. Here the measurement benefits from the high-precision luminosity 
measurement that ATLAS provides. With this method, $\sigmatot$ is given by the formula:
\begin{equation}\label{eq:totxs}
\sigmatot^{2} = \frac{16\pi(\hbar c)^2}{1+\rho^2} \left. \frac{\mathrm{d}\sigma_{\mathrm{el}}}{\mathrm{d}t}\right|_{t \rightarrow 0} \; ,  
\end{equation}
\end{linenomath*}
where $\rho$ represents a small correction arising from the ratio of the real to 
the imaginary part of the elastic-scattering amplitude in the forward direction and is taken from global 
model extrapolations \cite{PDG_2014}.

The first measurement of $\sigmatot$ at the LHC at 8 $\tev$ was performed by the TOTEM Collaboration \cite{TOTEM_8TeV} 
using a \textit{luminosity-independent }method and using data from the same LHC fill as ATLAS. At 7 $\tev$ 
measurements of $\sigmatot$ were provided by TOTEM \cite{TOTEM_first,TOTEM_second,TOTEM_lumindep} and ATLAS \cite{alfa_pub_7}.  
In a recent publication a measurement in the Coulomb--nuclear interference region at very small $t$ was also 
reported by TOTEM \cite{TOTEM_8TeV_1km}. 
The inelastic cross section $\sigmainel$ can either be derived from the total and elastic cross section measurements as 
in Refs. \cite{TOTEM_8TeV,TOTEM_first,TOTEM_second,TOTEM_lumindep,alfa_pub_7} at 7 and 8 $\tev$, or be 
determined directly from the measurement of the inelastic rate without exploiting the optical theorem.  
These measurements of $\sigmainel$ were performed 
at 7 $\tev$ by all LHC collaborations \cite{ATLAS_inel_7,CMS_inel_7,LHCb_inel_7,ALICE_inel_7,TOTEM_inel} and recently 
also at 13 $\tev$ by ATLAS \cite{ATLAS_inel_13}.

\section{Experimental setup}
\label{sec:exp}
The ATLAS detector is described in detail elsewhere \cite{atlas1}. The elastic-scattering data were recorded 
with the ALFA sub-detector (Absolute Luminosity For ATLAS) \cite{alfa_pub_7}. It consists of Roman Pot (RP) 
tracking-detector stations placed at distances of 237 m (inner station) and 241 m (outer station) on either side of the ATLAS interaction point (IP). 
Each station houses two vertically moveable scintillating fibre detectors which are inserted in RPs 
and positioned close to the beam for data taking. 
Each detector consists of 10 modules of scintillating fibres with 64 fibres on both the front and 
back sides of a titanium support plate. The fibres are arranged orthogonally in a $u$--$v$-geometry at $\pm45^\circ$ 
with respect to the $y$-axis. \footnote{ATLAS uses a right-handed coordinate 
system with its origin at the nominal IP in the centre of the detector and the $z$-axis along the 
beam pipe. The $x$-axis points from the IP to the centre of the LHC ring and the $y$-axis points upwards.}  
The spatial resolution of the detectors is 
about 35 $\mu$m. Elastic scattering events are recorded in two independent arms of the spectrometer. Arm 1 consists 
of two upper detectors at the left side and two lower detectors at the right side, and arm 2 consists inversely of 
two lower detectors at the left and two upper detectors at the right side. Events with reconstructed tracks in all four 
detectors of an arm are referred to as ``golden'' events \cite{alfa_pub_7}. 
The detectors are supplemented with trigger counters consisting of plain scintillator tiles. The detector geometry 
is illustrated in Figure~1 of Ref. \cite{alfa_pub_7}. All scintillation signals are detected by photomultipliers coupled 
to a compact assembly of front-end electronics including the MAROC chip \cite{MAROC1,MAROC2} 
for signal amplification and discrimination. The entire experimental setup is depicted in Figure~2 of Ref. \cite{alfa_pub_7}.

\section{Experimental method}
\label{sec:method}
\subsection{Measurement principle}
The data were recorded in a single run of the LHC with special beam optics \cite{Note90m,HBLumiday11} of 
$\betastar=90$ m. 
\footnote{The $\beta$-function determines the variation of 
the beam envelope around the ring and depends on the focusing properties of the magnetic lattice; its value at the 
IP is denoted by $\betastar$.} The same optics were 
used at $7 \tev$ \cite{alfa_pub_7} 
and result in a small beam divergence with parallel-to-point focusing in the vertical plane.
The four-momentum transfer $t$ is calculated from the scattering angle $\theta^\star$ and the beam momentum 
$p$ by:
\begin{linenomath*}
\begin{equation}
\label{eq:t-basic}
 -t = \left(\theta^\star \times p \right)^2 \,\,,
\end{equation}
where for the nominal beam momentum $p=3988 \pm 26 \gev$ is assumed \cite{Wenninger} and the scattering angle is 
calculated from the proton trajectories and beam optics parameters. The relevant beam optics parameters 
are incorporated in transport matrix elements which describe the particle trajectory from the interaction 
point through the magnetic lattice of the LHC to the RPs. Several methods were developed for the reconstruction 
of the scattering angle, as detailed in Ref. \cite{alfa_pub_7}. The {\em subtraction method} has the best resolution and 
is selected as the nominal method.    
It uses only the track positions ($w=\{x,y\}$)
and the matrix element $M_{12}=\sqrt{\beta \times \betastar}\sin\psi$,
where $\psi$ 
refers to the phase advance of the betatron function at the RP:
\begin{equation}
\label{eq:subtraction}
\theta^\star_w = \frac{w_\mathrm{A} - w_\mathrm{C}}{M_{12,\mathrm{A}} + M_{12,\mathrm{C}}} \; \; .
\end{equation}
\end{linenomath*}
Here A refers to the left side of the IP at positive $z$ and C refers to the right side at negative $z$. 
Three alternative methods are defined in detail in Ref. \cite{alfa_pub_7}. The {\em local angle method} uses only the $M_{22}$ 
matrix element and the track angle between the inner and outer detectors. 
The {\em local subtraction method} uses a combination of $M_{11}$ and $M_{12}$ matrix elements and both the local angle 
and track position.  
The {\em lattice method} also uses both track parameters and reconstructs the scattering angle by 
an inversion of the transport matrix. 
The alternative methods are used 
to impose constraints on the beam optics and to cross-check the subtraction method. 

\subsection{Data taking}
The low-luminosity, high $\betastar$ run had 
108 colliding bunches with about 7$\times$10$^{10}$ protons per bunch, 
but only 3 well-separated bunches of low emittance were selected for triggering. 
Precise positioning of the RPs is achieved with a 
beam-based alignment procedure which determines the position of the RPs with 
respect to the proton beams by monitoring the rate of the LHC beam-loss monitors during the RP insertion.
The data were collected with the RPs at a distance of approximately $7.5$ mm from the beam centre, corresponding 
to 9.5 times the vertical beam width. 
The beam centre and width monitored by LHC beam position monitors and the ATLAS beam-spot measurement \cite{BeamSpot} were 
found to be stable to within $10~\mu$m during the run. The beam emittance was derived from the width 
of the luminous region in conjunction with the beam optics. It was supplemented by direct measurement from ALFA in 
the vertical plane. The luminosity-weighted average of the emittance in the vertical plane was determined 
to be 1.6 $\mu$m for both beams and between 1.8 $\mu$m and 2.5 $\mu$m for beam 1 and beam 2 respectively 
in the horizontal plane. The emittance uncertainty is about $10\%$. 

To trigger on elastic-scattering events a coincidence was required between the A- and C-sides,
where on each side at least one trigger signal in a detector of the corresponding arm was required. 
The trigger efficiency 
was determined from a data stream recorded with looser conditions to be $99.9\%$ with negligible uncertainty. 
The dead-time fraction of the data acquisition system (DAQ) for the selected period was $0.4\%$. 

\subsection{Track reconstruction and alignment}
A well-reconstructed elastic-scattering event consists of local tracks from the 
proton trajectory in all four ALFA stations.    
The reconstruction method assumes that the protons pass through the fibre detector perpendicularly.  
The average multiplicity per 
detector is about 23 hits, where typically 18--19 are attributed to the proton trajectory 
while the remaining 4--5 hits are due to beam-related background, cross-talk and electronic noise. 
Tracks are reconstructed in several steps
from the overlap area of the hit fibres and several selections are
applied \cite{alfa_pub_7} in order to reject events with hadronic shower developments.

The precise detector positions  
with respect to the circulating beams are crucial inputs 
for the reconstruction of the proton kinematics.
First, the distance between the upper and the lower detectors is determined by the use of 
dedicated ALFA overlap detectors which allow simultaneous measurements of the same
particle in the upper and lower half of a station. 
Then, the detector positions are directly determined from the elastic-scattering data, 
using the fact that the high-$\betastar$ optics and the azimuthal symmetry of the scattering angle result 
in elastic hit patterns that have an ellipsoidal shape elongated in the vertical direction.
Three alignment parameters are determined for each detector: the horizontal and vertical offsets and the 
rotation angle around the beam axis. 
For the horizontal offset the centre of the $x$-distribution is taken and the rotation is obtained 
from a linear fit to a profile histogram of the $x$--$y$ correlation. 
The vertical offset 
is obtained from a comparison of the yields in the upper and lower detectors using the sliding window 
technique \cite{alfa_pub_7}. 
The above procedures provide an independent alignment of each ALFA station.
The vertical alignment parameters are in addition fine-tuned, exploiting the strong correlations between positions of tracks 
measured by different detectors in elastic events. First, the positions measured in one detector are extrapolated to the other 
detectors in the same arm using the ratio of the appropriate $M_{12}$ matrix elements.
Then, the extrapolated positions are compared to the corresponding measurements -- the average distance gives information 
about residual misalignments. The residuals obtained for all pairs of detectors are combined with the vertical offset and 
distance measurements in a global $\chi^2$ fit, resulting in the final alignment parameters.

\section{Model for elastic scattering simulation}
\label{sec:thmc}

Several parameterizations are available \cite{Bethe,WestAndYennie,Bourrely_Spin,Cahn,menon_silva,
Block_and_Cahn_curvature,KFK,Selyugin,PhillipsAndBarger,Fagundes,BourrelyAndSoffer} for the differential elastic $pp$ cross section.  
A conventional approach is adopted here by taking the following simplified formulae:
\begin{linenomath*}
\begin{eqnarray}
\label{eq:elamplitudes}
\frac{\mathrm{d}\sigma}{\mathrm{d}t} & = & \frac{1}{16\pi}\left|f_{\mathrm{N}}(t) + f_{\mathrm{C}}(t)\mathrm{e}^{\mathrm{i}\alpha\phi(t)}\right|^2 \; \; , \\
 f_{\mathrm{C}}(t) & = & -8\pi\alpha\hbar c\frac{G^2(t)}{|t|} \;\; , \\
 f_{\mathrm{N}}(t) & = & \left(\rho + \mathrm{i}\right)\frac{\sigmatot}{\hbar c}\mathrm{e}^{-B|t|/2} \;\; , 
\end{eqnarray}
\end{linenomath*}
where $G$ is the electric form factor of the proton, $B$ the nuclear slope, $f_{\mathrm{C}}$ the Coulomb amplitude 
and $f_{\mathrm{N}}$ the nuclear amplitude with $\phi$ their relative phase shift. 
The value of $\rho=\mathrm{Re}(f_{\mathrm{el}}) / \mathrm{Im}(f_{\mathrm{el}}) = 0.1362 \pm 0.0034$ is taken 
from a global fit to lower-energy data \cite{PDG_2014} and parameterizations for $G$ and $\phi$ are given 
in Ref. \cite{alfa_pub_7}. This expression is used to fit the data and extract 
$\sigmatot$ and $B$. 

Monte Carlo simulation of elastic-scattering events is performed with PYTHIA8~\cite{PYTHIA,PYTHIA6} version 8.186 
with a $t$-spectrum generated according to Eq.~(\ref{eq:elamplitudes}). 
The simulation is used to calculate acceptance and unfolding corrections. 
In the simulation the angular divergence of beams at the IP and the spread of the production vertex 
are set to the measured values. 
Elastically scattered protons are transported from the interaction 
point to the RPs nominally by means of the transport matrix. For studies of systematic uncertainties this was also done by the tracking 
module of the MadX~\cite{madx} beam optics calculation program. A fast parameterization of the detector response is 
used in the simulation and tuned to reproduce the measured difference in position between the outer detectors and 
their position as extrapolated from the inner detectors. 

\section{Data analysis}
\label{sec:analysis}
\subsection{Event selection}
\begin{figure}[h!]
  \centering
   \includegraphics[width=0.48\textwidth]{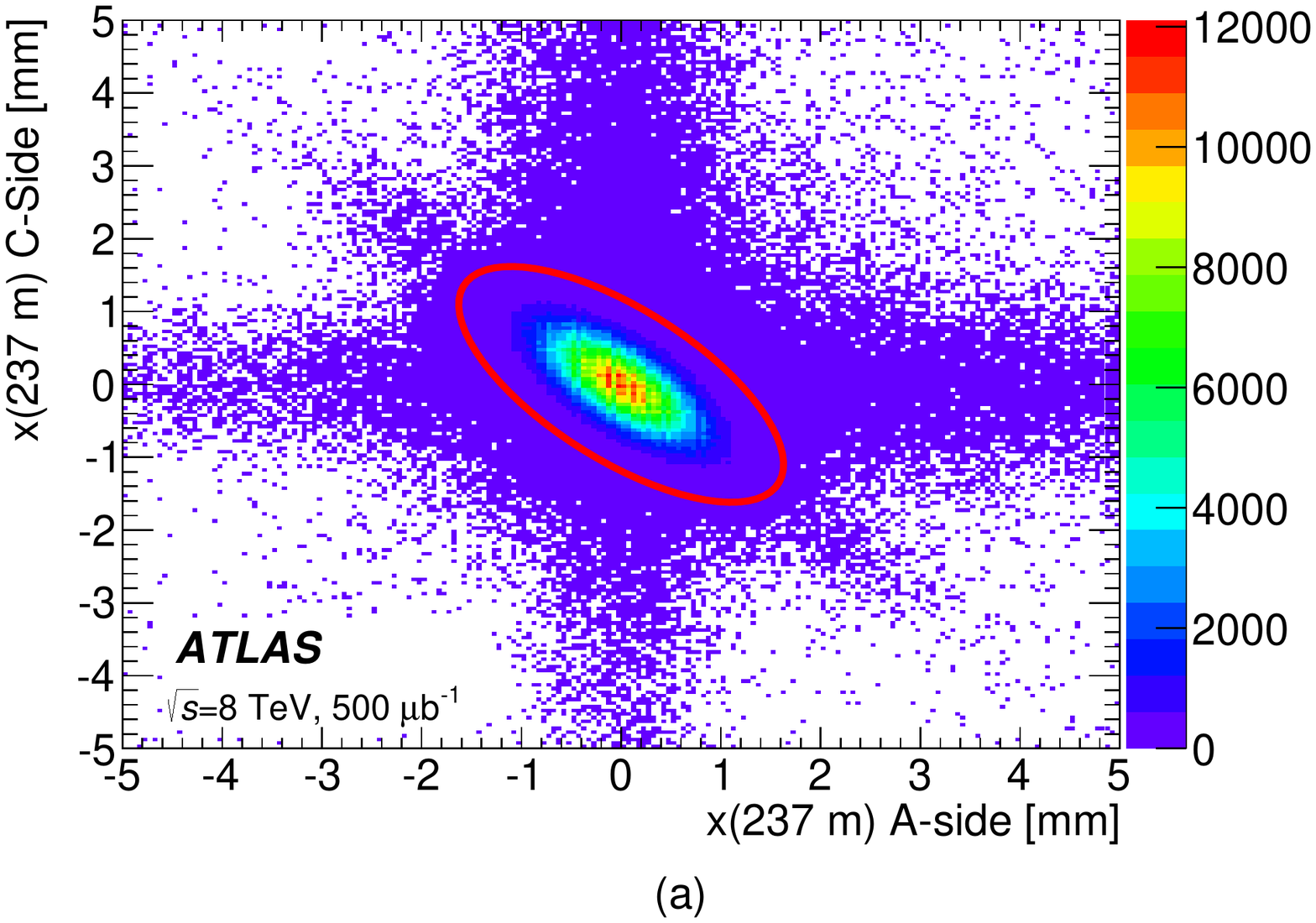} 
   \includegraphics[width=0.48\textwidth]{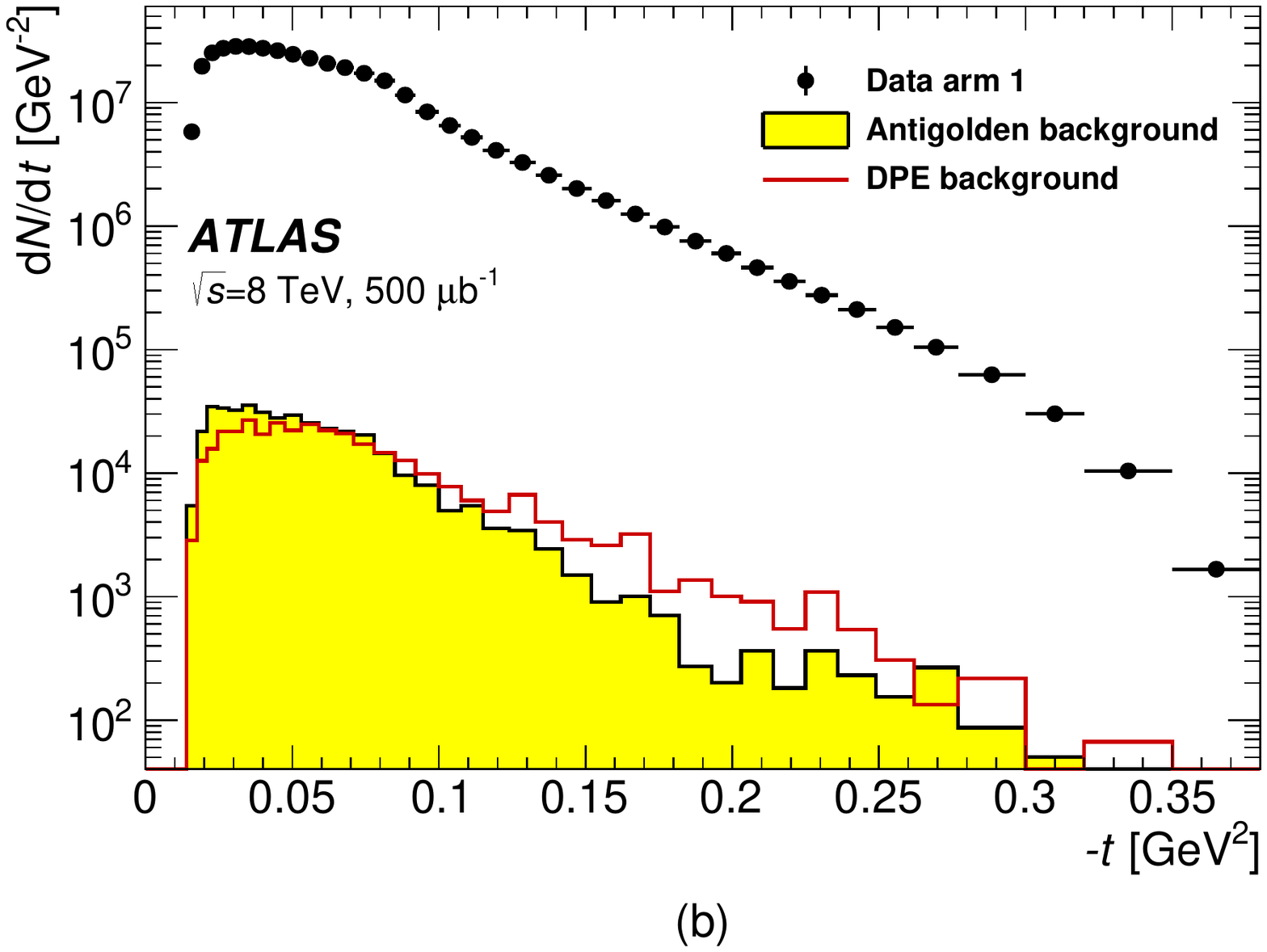}
\label{fig:background_antigolden1}
     \caption{(a) The correlation between the horizontal coordinates on the A- and C-sides.  
Elastic-scattering candidates after data quality, trigger and bunch selection but before 
acceptance and background rejection cuts are shown. Identified elastic events are required to lie inside the ellipse. 
\noindent (b) The distribution $\mathrm{d}N/\mathrm{d}t$, before corrections, as a function of $t$ in arm 1 compared
  to the background spectrum determined using anti-golden events. 
  The results of a simulation of the DPE background is also shown for comparison.} 
\label{fig:evsel}
\end{figure}

Events are required to pass the trigger conditions for elastic-scattering events 
and have a reconstructed track in all four detectors of an arm in the golden topology. 
The fiducial volume 
is defined by cuts on the vertical coordinate of the reconstructed track, which 
is required to be at least 90 $\mu$m from the detector edge near the beam and at least 1 mm away from 
the shadow of the beam screen, in each of the four detectors. 
\footnote{The beam screen is a protection element of the quadrupoles, which limits the acceptance of the detector at large $|y|$.}  
The values of cuts are chosen to obtain good agreement between data and simulation in the position distributions. 
The back-to-back topology of elastic events is further exploited to clean the sample by imposing  
cuts on the left-right acollinearity. The difference between the absolute value of the vertical coordinate at the A- and 
C-side is requested to be below 3 mm. For the horizontal coordinate the correlation of the A- and C-sides is used.  
Events are selected inside an ellipse with half-axis values of $3.5 \sigma$ of the resolution determined by simulation, 
as illustrated in Figure~\ref{fig:evsel}(a).  
Elastic events are concentrated inside a narrow ellipse with negative slope, whereas the 
beam-halo background appears in broad uncorrelated bands.
The most efficient selection against background is obtained from  
the correlation between the position in the horizontal plane and the local angle between two stations,  
where events on either side are again required to be inside an ellipse of $3.5\sigma$ width. 
From an initial sample of 4.2 million elastic candidates, 3.8 million golden elastic events were selected after all cuts.  
The $t$-spectrum, before corrections, for selected elastic events in one arm is shown in Figure~\ref{fig:evsel}(b). 

\subsection{Background estimate}
A small fraction of the events inside the selected elliptical area shown in
Figure~\ref{fig:evsel}(a) are expected to be background, predominantly originating from double-Pomeron exchange (DPE)
according to simulations based on the MBR model~\cite{MBR}. The background is estimated 
with a data-driven method \cite{alfa_pub_7} using events in the ``anti-golden'' topology 
with two tracks in both upper or both lower detectors at the A- and C-sides. This 
sample is free of signal and yields an estimate of background in the elastic sample with the golden topology. 
The shape of the $t$-spectrum for background events is obtained by flipping the sign of the vertical coordinate 
on either side. The resulting background distribution is shown in Figure~\ref{fig:evsel}(b). In total 4400 background 
events are estimated to be in the selected sample, corresponding 
to a fraction of $0.12\%$ of the selected events. The systematic uncertainty is about 50\%, as derived in 
Ref.~\cite{alfa_pub_7} from a comparison of different methods.

\subsection{Reconstruction efficiency}
The rate of elastic-scattering events is corrected for reconstruction inefficiencies.  
These events may not be reconstructed 
when protons or halo particles interact with the stations or detectors, causing a shower to develop and   
resulting in high fibre hit multiplicities. 
This correction is called the event reconstruction efficiency and is given by
\begin{linenomath*}
\begin{equation}\label{eq:eff2}
\varepsilon_{\text{rec}} = \frac{N_{\text{reco}}}{N_{\text{reco}}+N_{\text{fail}}} \; ,
\end{equation}
\end{linenomath*}
for each arm where $N_{\text{reco}}$ is the number of reconstructed events and $N_{\text{fail}}$ the number of events for 
which the reconstruction failed because a shower developed. The sample of failed events is split into 
different categories depending 
on the number of detectors with reconstruction failures, because the event background is different for each category.  
The fraction of elastic events in the subsample where one out of four detectors failed to reconstruct a track 
is above 99\%, 
whereas this fraction is 95\% for the subsample where two detectors failed to reconstruct a track on one side. 
The event yields in the 
different categories are 
calculated with a data-driven method, for which the details are given in Ref. \cite{alfa_pub_7}. 
The background fraction in the case with only two detectors with reconstructed tracks is estimated with background 
templates of the $x$ distribution, obtained from data by selecting single diffractive events.  
In the case of a successful track reconstruction in three detectors, 
where a good $t$-measurement 
is still possible, the partial reconstruction efficiency was verified to be independent of $t$, which is then also 
assumed for the other categories. 
Events falling outside the acceptance, but faking a signal through shower development, were eliminated from the 
reconstruction efficiency calculation by applying another template analysis using the $y$ distribution obtained 
from golden elastic events. 

The event reconstruction efficiencies in arm 1 and arm 2 are determined to be
$\varepsilon_{\text{rec,1}} = 0.9050 ~ \pm 0.0003 \stat ~ \pm ~ 0.0034 \syst$ and 
$\varepsilon_{\text{rec,2}} = 0.8883 ~ \pm ~ 0.0003 \stat ~ \pm ~ 0.0045 \syst$, respectively. 
The lower reconstruction efficiency in arm 2 originates from a different amount of material which 
induces a higher probability of shower development. 
The systematic uncertainty is estimated   
by a variation of the selection criteria and templates, 
as described in Ref. \cite{alfa_pub_7}.

\subsection{Beam optics}
The precision of the $t$-reconstruction depends on knowledge of the transport matrix elements. 
A data-driven method was developed \cite{alfa_pub_7} to tune the relevant matrix elements using 
constraints on the beam optics derived from measured correlations in the ALFA data. These constraints 
are incorporated in a fit of the strength of the inner triplet quadrupole magnets Q1 and Q3, which 
yields an effective beam optics used in the simulation. The values of the constraints are compatible with those  
published in Ref. \cite{alfa_pub_7} within $15\%$ and the resulting magnet strength offsets are in good 
agreement with the values found at $7~ \tev$.

\subsection{Acceptance and unfolding}
\begin{figure}[h!]
  \centering
  \includegraphics[width=0.5\textwidth]{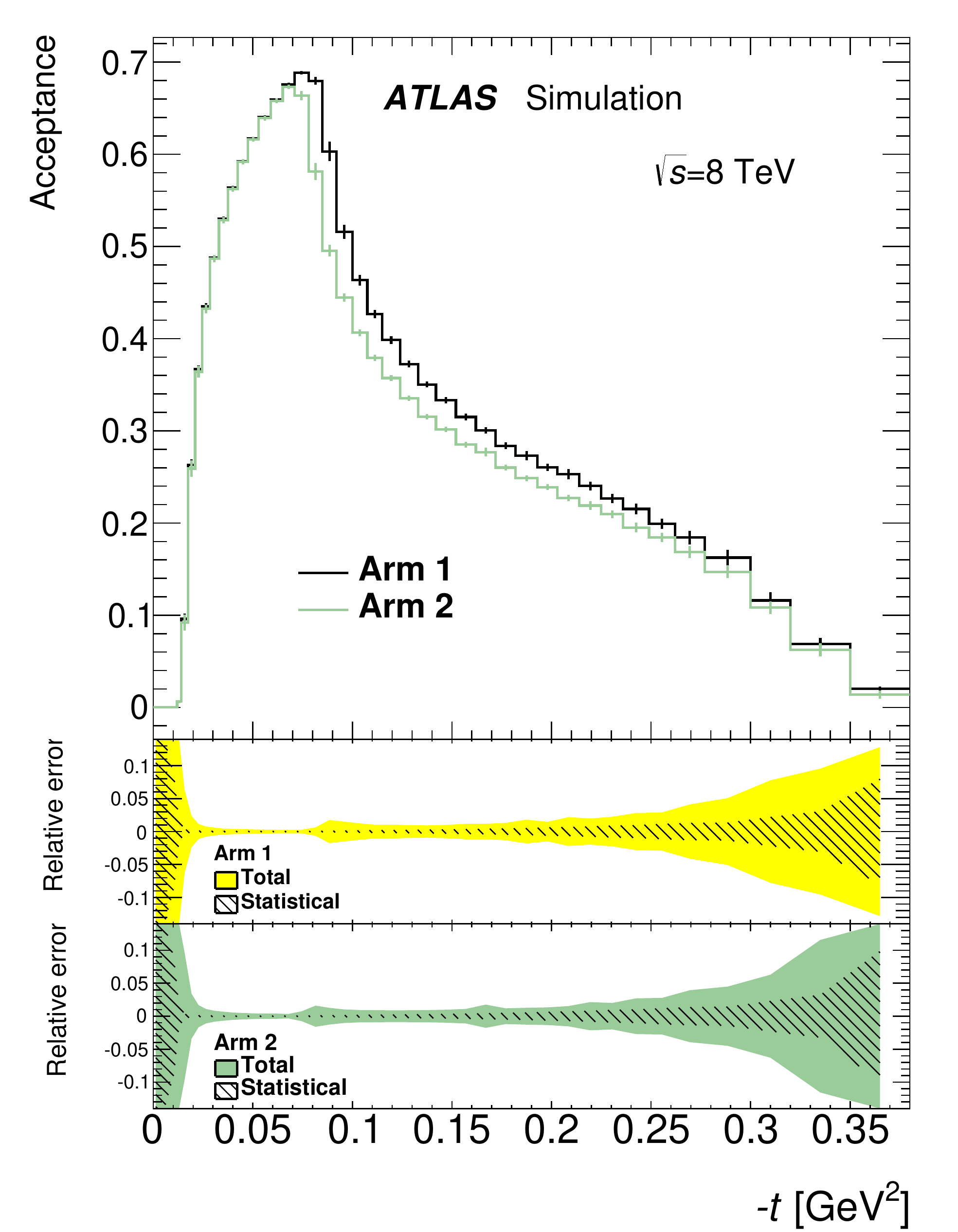}
  \caption{The acceptance as a function of the true value of $t$ for each arm with total uncertainties shown 
as error bars. The lower panels show relative total and statistical uncertainties.} 
  \label{fig:acceptance}
\end{figure}

The acceptance is defined as the ratio of events passing all geometrical 
and fiducial acceptance cuts to all generated events, and is 
calculated as a function of $t$. The form of the acceptance curve as shown in Figure~\ref{fig:acceptance} 
results from the different contributions 
of the vertical and horizontal scattering angles to the value of $t$ and the impact of the fiducial volume cuts 
on these contributions. In particular, the position of the peak depends on the cut at large $|y|$ at 
the beam screen, which is slightly different for the two arms. The rise of the acceptance at small 
$t$ is different in the two arms because of different detector distances, between 8 and 8.4 mm, to the beam. 

The measured $t$-spectrum is affected by detector resolution and beam divergence 
effects, which are corrected with an unfolding procedure. The $t$-resolution of the subtraction method  
is about $10\%$ at small $t$ and $3\%$ at large $t$. The alternative methods have 
a $t$-resolution which is a factor of 2--3 worse \cite{alfa_pub_7}.  
The background-subtracted distributions 
in each arm are corrected for migration effects using an iterative, dynamically stabilized, unfolding 
method~\cite{IDS}, which is based on a simulated transition matrix describing the resolution-induced 
migration between bins of the $t$-spectrum. The corrections induced by the unfolding are small ($<2\%$) for 
the subtraction method except at small $t$ where they rise to $30\%$. 
For the other methods the corrections are generally $t$-dependent and increase to $50\%$ 
at large $t$. 
\subsection{Luminosity}
The ATLAS luminosity measurement at high luminosity ($L>10^{33}$cm$^{-2}$s$^{-1}$) is described in detail in 
Ref.~\cite{LumiPaper2012}.
Unlike that measurement, the run in this analysis had an instantaneous luminosity 
$L \sim0.05\cdot10^{30}$cm$^{-2}$s$^{-1}$, about five orders of magnitude lower. 
Only three bunches were present in this run, whereas more than a thousand bunches are common at high luminosity. 
The average number of interactions per bunch-crossing (pile-up) in this sample is $\mu\sim0.1$, 
which is also low compared to the values of $\mu$=10--40 reached routinely in normal conditions. 
At such low values of the luminosity,   
some of the standard algorithms are unusable due to lack of sensitivity.  
On the other hand, an additional method based on vertex counting in the inner 
detector (ID) can be exploited, which is most effective at low pile-up. 
Another consequence of the low luminosity is the relative importance of the background sources: 
the beam--gas contribution, normally negligible, can become 
comparable with the collision rate, while the “afterglow” background (see Ref.~\cite{LumiPaper2012})  
becomes conversely less important, due to the small number of colliding bunches. 

In 2012, the beam conditions monitor (BCM) was used as the baseline detector for luminosity measurements.  
It consists of diamond-sensor detectors placed
on both sides of the IP. It measures the luminosity using an event-counting 
method based on the requirement of having activity in either side (BCM\_EventOR). 
LUCID (LUminosity measurement with a Cherenkov Integrating Detector) is also located 
on both sides of the IP and uses the same algorithm to measure the luminosity (LUCID\_EventOR). 
A third method for measuring the per-bunch luminosity is provided by the ID.
Tracks are reconstructed requiring at least nine hits and no missing hits along the track trajectory, and a 
transverse momentum $p_{\mathrm{T}}>$ 900 $\MeV$. Then, at least five selected tracks are required to form a primary vertex (VTX5). 
The number of primary vertices per event is proportional to the luminosity and provides an
independent method with respect to LUCID and BCM.

The absolute luminosity scale of each algorithm was calibrated by the van der Meer (vdM) method ~\cite{svdm} 
at an intermediate luminosity regime ($L\sim10^{30}$cm$^{-2}$s$^{-1}$). 
The treatment of both afterglow and beam--gas background is described in detail in 
Ref.~\cite{LumiPaper2012}. The first is evaluated by measuring the detector activity 
in unfilled bunches preceding the colliding bunches, while the second is estimated from the so-called unpaired bunches, 
in which bunches in only one of the two beams are filled and no beam--beam collisions occur. 
In the high-$\betastar$ run and for BCM\_EventOR, the afterglow background is evaluated to be 0.05\% and   
the beam--gas contribution is 0.4\%. 

BCM\_EventOR was chosen as the baseline algorithm for the luminosity determination, 
whereas the LUCID\_EventOR and VTX5 methods are only used for the evaluation of systematic uncertainties. 
It proved to be the most stable, both by comparing the various vdM calibration
sessions performed during the year and by studying its long-term behaviour at high luminosity. This choice 
also ensures maximum compatibility with the high-luminosity case.
By comparing the LUCID\_EventOR and VTX5 results with BCM\_EventOR, a maximum
difference of 0.3\% is found. No change 
of this difference with time, or equivalently $\mu$, is observed.

The following contributions to the systematic uncertainty of the luminosity determination are considered:
\begin{itemize}
\item The absolute luminosity scale, common to all algorithms, is determined by the vdM method. Its uncertainty of 1.2\%   
is dominated by the beam conditions. This uncertainty is fully correlated between low- and high-luminosity 
data sets~\cite{LumiPaper2012}.  
\item The BCM calibration stability between the high-$\betastar$ run and the vdM session is estimated to be 0.8\% 
by comparing with the VTX5 method among the various vdM scans. 
\item The afterglow background uncertainty is conservatively taken as 100\% of the afterglow level itself, which leads to an 
uncertainty of 0.05\% in the luminosity.
\item The beam--gas background uncertainty is obtained using LUCID by comparing the difference in the off-time activity 
(i.e. produced by beam--gas interactions and not by collisions at the IP) between the colliding and the unpaired bunches. 
It is estimated to be 0.3\%. 
\end{itemize}
The total systematic uncertainty is therefore 1.5\%. 
The final integrated luminosity is measured to be ${L_\mathrm{{int}}} = 496.3 ~ \pm ~ 0.3 ~ (\mathrm{stat.}) ~ \pm ~ 7.3 ~ (\mathrm{syst.})~\mu$b$^{-1}$.

\section{Results}
\label{sec:result}
\subsection{Elastic cross section}
The differential elastic cross section in a given bin $t_i$ is calculated from the following formula:
\begin{linenomath*}
\begin{equation}\label{eq:cross-section}
\frac{\mathrm{d}\sigmael}{\mathrm{d}t_i} = \frac{1}{\Delta t_i}\times \frac{{\cal M}^{-1}[N_i - B_i]}{A_i  \times \epsilon^{\mathrm{reco}} \times \epsilon^{\mathrm{trig}} \times \epsilon^{\mathrm{DAQ}}  \times L_{\mathrm{int}} }\; \; , 
\end{equation}
\end{linenomath*}
where $\Delta t_i$ is the width of the bins in $t$, ${\cal M}^{-1}$ symbolizes the unfolding procedure applied to the 
background-subtracted number of events $N_i - B_i$, $A_i$ is the acceptance,  
$\epsilon^{\mathrm{reco}}$ is the event reconstruction efficiency, $\epsilon^{\mathrm{trig}}$ is the trigger efficiency, 
$\epsilon^{\mathrm{DAQ}}$ is the dead-time correction and $L_{\mathrm{int}}$ is the integrated luminosity. The binning 
in $t$ is chosen to yield a purity above $50\%$, which corresponds to 1.5 times the resolution at small $t$.  
It is enlarged at large $t$ in order to account for the lower number of events. The numerical values for the resulting 
differential elastic cross section are given in Table~\ref{tab:differential_elastic_crosssection}. 
\begin{table}[!htb]
\scriptsize
  \begin{center}  
    \begin{tabular}{S[table-format=1.4] S[table-format=1.4] S[table-format=1.4] S[table-format=3.2] S[table-format=2.2] S[table-format=2.2] S[table-format=2.2]}
      \hline \hline
{Low $|t|$ edge}  & {High $|t|$ edge} & {Central $|t|$} & {$\mathrm{d}\sigma_{\mathrm{el}}/\mathrm{d}t$} & {Stat. uncert.}  & {Syst. uncert.} & {Total uncert.}\\
{[\GeV$^2$]}   & {[\GeV$^{2}$]} &  {[\GeV$^{2}$]} & {[mb/\GeV$^2$]} & {[mb/\GeV$^2$]} & {[mb/\GeV$^2$]} & {[mb/\GeV$^2$]} \\ \hline
0.0090 & 0.0120 & 0.0105 & 387 & 29 & 14 & 32 \\
0.0120 & 0.0140 & 0.0130 & 370 & 5.6 & 12 & 13 \\
0.0140 & 0.0175 & 0.0157 & 352.3 & 1.4 & 8.7 & 8.9 \\
0.0175 & 0.0210 & 0.0192 & 329.8 & 0.8 & 6.5 & 6.5 \\
0.0210 & 0.0245 & 0.0227 & 306.9 & 0.6 & 5.7 & 5.8 \\
0.0245 & 0.0285 & 0.0265 & 284.6 & 0.5 & 5.0 & 5.1 \\
0.0285 & 0.0330 & 0.0307 & 261.7 & 0.4 & 4.6 & 4.6 \\
0.0330 & 0.0375 & 0.0352 & 239.3 & 0.4 & 4.1 & 4.1 \\
0.0375 & 0.0425 & 0.0400 & 218.0 & 0.4 & 3.6 & 3.6 \\
0.0425 & 0.0475 & 0.0450 & 197.3 & 0.3 & 3.3 & 3.3 \\
0.0475 & 0.0530 & 0.0502 & 178.0 & 0.3 & 3.0 & 3.0 \\
0.0530 & 0.0590 & 0.0559 & 158.8 & 0.2 & 2.7 & 2.7 \\
0.0590 & 0.0650 & 0.0619 & 141.1 & 0.2 & 2.4 & 2.4 \\
0.0650 & 0.0710 & 0.0679 & 126.0 & 0.2 & 2.2 & 2.2 \\
0.0710 & 0.0780 & 0.0744 & 111.1 & 0.2 & 2.0 & 2.0 \\
0.0780 & 0.0850 & 0.0814 & 96.8 & 0.2 & 2.0 & 2.0 \\
0.0850 & 0.0920 & 0.0884 & 84.7 & 0.2 & 1.7 & 1.7 \\
0.0920 & 0.1000 & 0.0959 & 72.9 & 0.2 & 1.6 & 1.6 \\
0.1000 & 0.1075 & 0.1037 & 62.7 & 0.2 & 1.5 & 1.5 \\
0.1075 & 0.1150 & 0.1112 & 54.1 & 0.2 & 1.3 & 1.4 \\
0.1150 & 0.1240 & 0.1194 & 46.11 & 0.14 & 1.13 & 1.13 \\
0.1240 & 0.1330 & 0.1284 & 38.76 & 0.14 & 1.0 & 1.01 \\
0.1330 & 0.1420 & 0.1374 & 32.60 & 0.12 & 0.92 & 0.93 \\
0.1420 & 0.1520 & 0.1468 & 27.10 & 0.11 & 0.82 & 0.83 \\
0.1520 & 0.1620 & 0.1568 & 22.48 & 0.11 & 0.74 & 0.74 \\
0.1620 & 0.1720 & 0.1668 & 18.48 & 0.10 & 0.68 & 0.68 \\
0.1720 & 0.1820 & 0.1768 & 15.25 & 0.09 & 0.67 & 0.68 \\
0.1820 & 0.1930 & 0.1873 & 12.36 & 0.08 & 0.57 & 0.58 \\
0.1930 & 0.2030 & 0.1978 & 10.08 & 0.08 & 0.48 & 0.48 \\
0.2030 & 0.2140 & 0.2083 & 8.20 & 0.07 & 0.43 & 0.43 \\
0.2140 & 0.2250 & 0.2193 & 6.58 & 0.06 & 0.33 & 0.33 \\
0.2250 & 0.2360 & 0.2303 & 5.34 & 0.06 & 0.27 & 0.28 \\
0.2360 & 0.2490 & 0.2422 & 4.28 & 0.05 & 0.24 & 0.24 \\
0.2490 & 0.2620 & 0.2552 & 3.30 & 0.05 & 0.22 & 0.23 \\
0.2620 & 0.2770 & 0.2691 & 2.47 & 0.04 & 0.18 & 0.18 \\
0.2770 & 0.3000 & 0.2877 & 1.69 & 0.03 & 0.14 & 0.14 \\
0.3000 & 0.3200 & 0.3094 & 1.06 & 0.03 & 0.10 & 0.1 \\
0.3200 & 0.3500 & 0.3335 & 0.62 & 0.02 & 0.08 & 0.08 \\
0.3500 & 0.3800 & 0.3635 & 0.36 & 0.04 & 0.04 & 0.05 \\ \hline
\end{tabular}
  \caption{The measured values of the differential elastic cross section with statistical and systematic uncertainties. 
The central $t$-values in each bin are calculated from simulation, in which a slope parameter of $B = 19.7  \GeV^{-2}$ is used.}
  \label{tab:differential_elastic_crosssection} 
  \end{center}
\end{table}

The experimental systematic uncertainties are derived according to the methods detailed in Ref. \cite{alfa_pub_7}
as follows:
\begin{itemize}
 \item The value of the beam momentum used in the $t$-reconstruction (Eq.~(\ref{eq:t-basic})) and in the simulation is 
 varied by $0.65\%$, as recommended in Ref.~\cite{Wenninger}.
 \item The uncertainty in the luminosity of $1.5\%$ is applied to the cross-section normalization.
 \item The event reconstruction efficiency is varied by its uncertainty of about $0.5\%$ and the uncertainty in the tracking efficiency 
 is estimated by varying the reconstruction criteria.
 \item The uncertainties originating from the effective beam optics are calculated from variations of the 
 optics constraints, of the strength of the quadrupoles not adjusted in the fit, and of the quadrupole alignment 
 constants. Additional uncertainties are related to the error of the optics fit, 
 to the beam transport scheme used in the simulation, and to the impact from a residual 
 beam crossing angle assumed to vary within its uncertainty of  
 $\pm$10 $\mu$rad. 
 \item The uncertainties from the alignment of the ALFA detectors are evaluated by varying the correction
constants for horizontal and vertical offsets as well as the rotation
within their uncertainties as determined from variations of the
alignment procedures, and by taking the difference between different optimization configurations for the vertical alignment parameters. 
 \item The background normalization uncertainty of $50\%$ is applied in the background subtraction and 
 the background shape 
 is varied by inverting the sign of different detector combinations.
 \item The detector resolution values in the fast simulation are replaced by estimates from GEANT4~\cite{GEANT41,GEANT42} and test-beam 
 measurements, and a $y$-dependent resolution is used instead of a constant value.
 \item The value of the nuclear slope in the simulation is varied around the nominal value of 
 $19.7 \gev^{-2}$ by $\pm 1 \gev^{-2}$, corresponding to about five times the uncertainty of the measured $B$ value. 
 \item The beam emittance value in the simulation is varied by its uncertainty of about $7\%$. Additionally, the ratio of 
 the emittance in beam 1 to the emittance in beam 2, which are measured by wire scans after injection only, is set to unity. 
 \item The intrinsic unfolding uncertainty is estimated from a data-driven closure test.
\end{itemize}
The main sources of systematic uncertainty are the beam momentum uncertainty and the luminosity uncertainty. For each 
systematic uncertainty source the shift of the cross-section value in each $t$-bin is recorded. 
The most important shifts are shown in Figure~\ref{fig:tfit}(a).

\subsection{Total cross section}
A profile fit~\cite{profile} is used to determine $\sigmatot$. 
It includes statistical and systematic uncertainties 
and their correlations across the $t$-spectrum. For each shift due to a systematic uncertainty a nuisance parameter is fitted in 
a procedure described in Ref.~\cite{alfa_pub_7}.  

\begin{figure}[h!]
  \centering
   \includegraphics[width=0.48\textwidth]{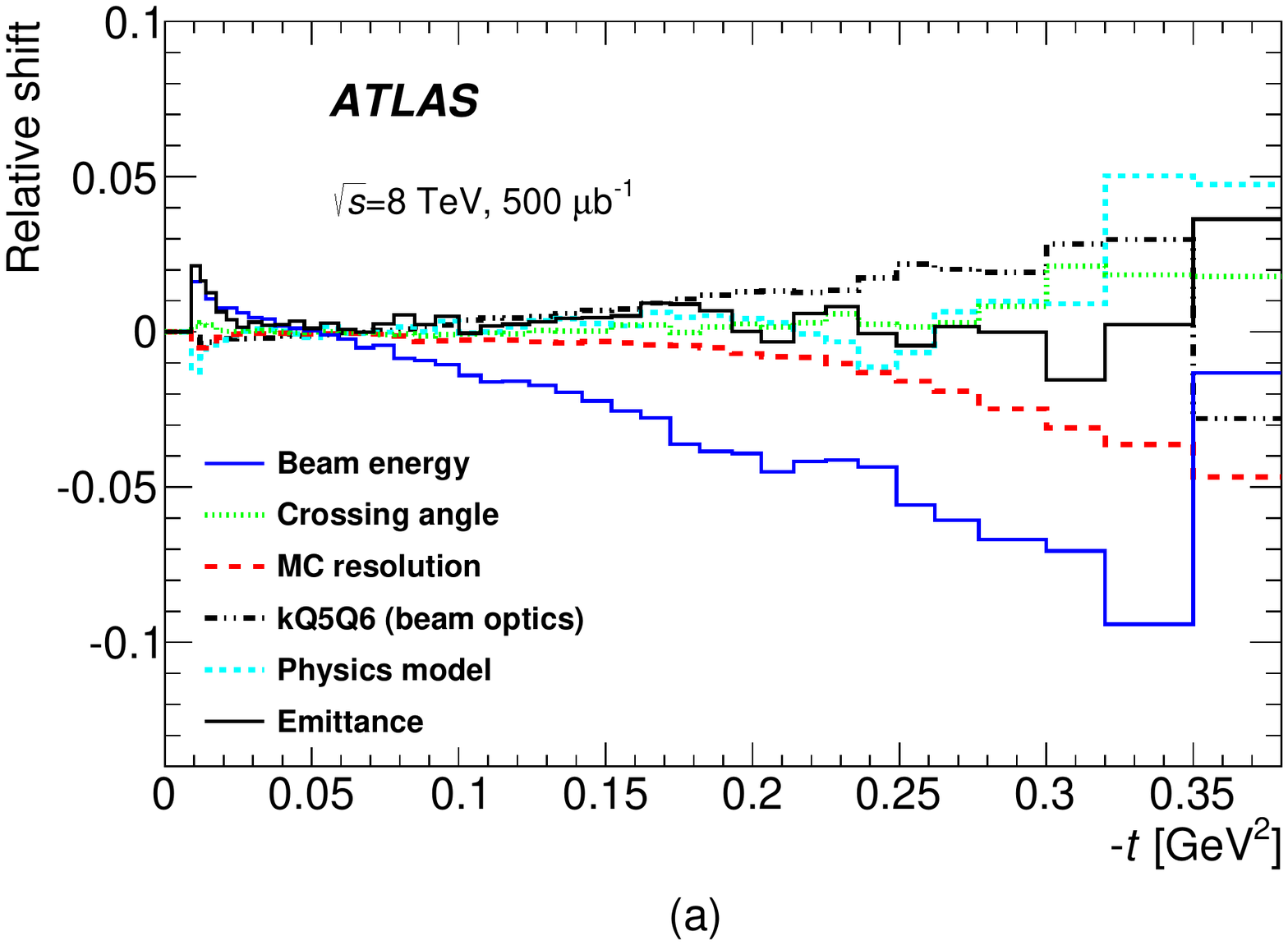}
   \includegraphics[width=0.48\textwidth]{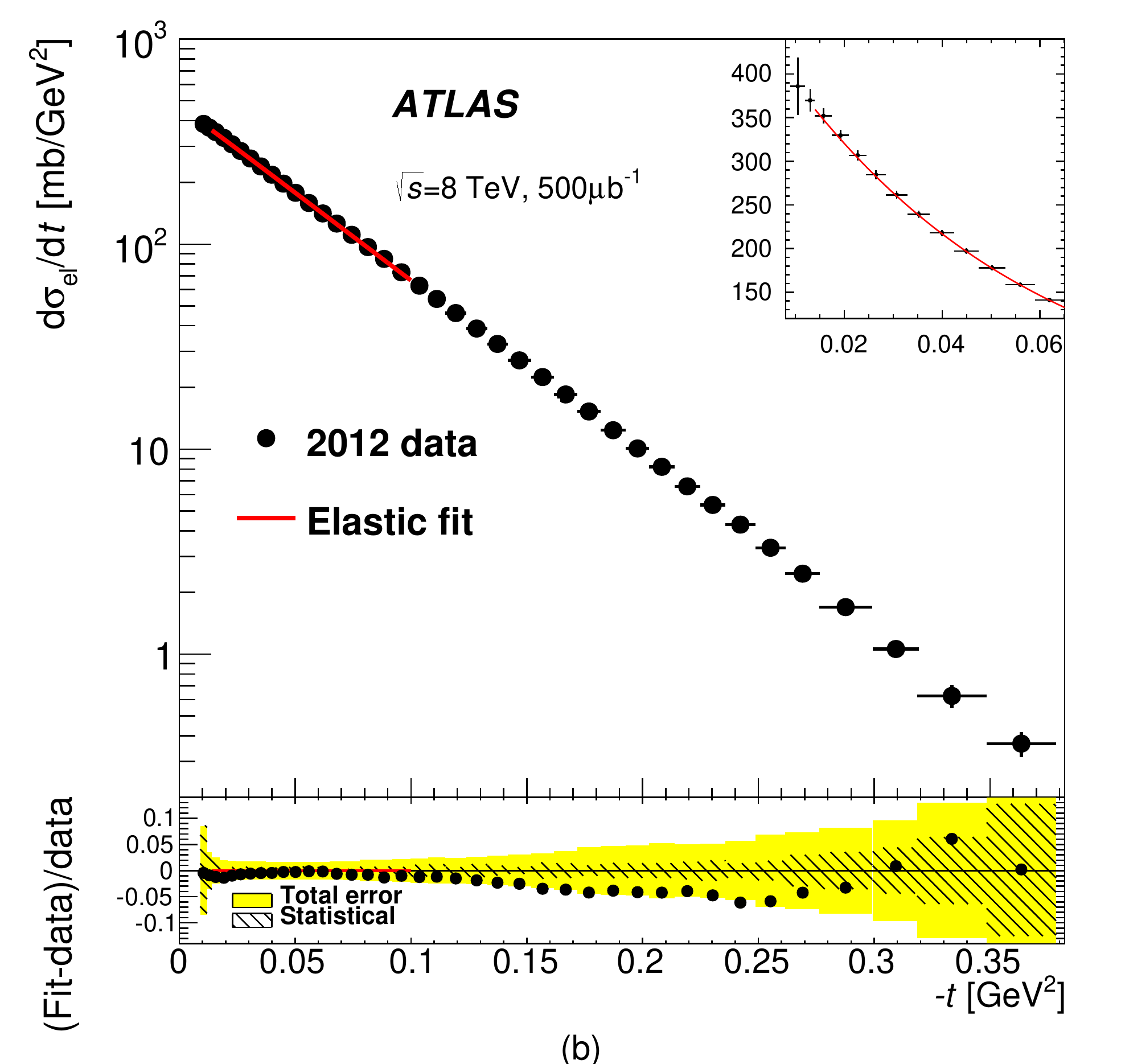} 
  \caption{(a) Relative shifts in the differential elastic cross section as a function of $t$ for 
  selected systematic uncertainty sources. Shown are the uncertainties related to the beam energy, to the crossing angle, to the 
  modelling of the detector resolution in the simulation (MC resolution), to the beam optics (kQ5Q6, magnet strength), to the value 
  of $B$ in the simulation (Physics model) and to the emittance.
  (b) The fit of the theoretical prediction to the
 differential elastic cross section with $\sigmatot$ and $B$ as free parameters.
In the lower plot the points represent the relative difference between fit and data, 
 the yellow area represents the total experimental uncertainty and the hatched area the statistical component. 
 The red line indicates the fit range; the fit result is extrapolated in the lower plot outside the fit range. 
 The upper right insert shows a zoom of the data and fit at small $t$.
} 
  \label{fig:tfit}
\end{figure}

The theoretical prediction of Eq.~(\ref{eq:elamplitudes}) including the Coulomb and interference terms is fitted to the data 
to extract $\sigmatot$ and $B$ alongside the nuisance parameters, as shown in Figure~\ref{fig:tfit}(b).
The fit range is chosen to be from $-t = 0.014 \GeV^2$ to $-t = 0.1 \GeV^2$, 
where the lower bound is set by requiring the acceptance to exceed $10\%$ and the upper 
bound is chosen to exclude the large-$t$ region where theoretical models predict 
deviations from a single exponential function \cite{per_theory}. 
The fit yields 
$\sigmatot = 96.07\pm0.86~\mbox{mb}$ and $ B = 19.74\pm0.17 \GeV^{-2}$ with $\chi^2/N_{\mbox{dof}}=17.8/14$  
and the uncertainties include all statistical and experimental systematic contributions. The most important 
uncertainty component is the luminosity error for $\sigmatot$ and the beam energy error for $B$.  
Additional uncertainties arising from the extrapolation $t\rightarrow 0$ are  
estimated from a variation of the upper end of the fit range respectively up to $-t=0.152 \GeV^2$ 
and up to $-t=0.065 \GeV^2$, and from a variation of the lower end,  
i.e. from $-t=0.009 \GeV^2$ to $-t=0.0245 \GeV^2$. Further theoretical uncertainties considered include:  
a variation of the $\rho$-parameter in Eq. (\ref{eq:OpticalTheorem}) by $\pm 0.0034$; the replacement of the dipole parameterization by a double-dipole 
parameterization~\cite{A1} for the proton 
electric form factor; the replacement of the Coulomb phase from West and Yennie \cite{WestAndYennie}
by parameterizations from Refs.~\cite{Cahn,KFK}; the inclusion of a term related to the magnetic moment 
of the proton in the Coulomb amplitude~\cite{Bourrely_Spin}. The dominant extrapolation uncertainty is induced 
by the fit range variation. The final results for $\sigmatot$ and $B$ are: 
\begin{eqnarray}\label{eq:profilefit}
\sigmatot & = &  \mbox{96.07} \; \pm \mbox{0.18} \; (\mbox{stat.}) \pm \mbox{0.85} \; (\mbox{exp.})  \pm \mbox{0.31} \; (\mbox{extr.})  \; \mbox{mb} \; , \\
B & = & 19.74 \; \pm \mbox{0.05} \; (\mbox{stat.}) \pm \mbox{0.16} \; (\mbox{exp.})  \pm \mbox{0.15} \; (\mbox{extr.}) \;  \GeV^{-2} \; .  
\end{eqnarray}
A summary of the results for $\sigmatot$ from four different $t$-reconstruction methods 
is given in Table~\ref{tab:sigma_tot_syst_summary}. 
\begin{table}
  \begin{center}
    \begin{tabular}{l|S[table-format=2.2] | S[table-format=2.2] S[table-format=2.2] S[table-format=2.2]}
      \hline \hline
            & \multicolumn{4}{c}{$\sigmatot$ [mb]} \\          
	   & {Subtraction} & {Local angle} & {Lattice} & {Local subtraction} \\ \hline 
Total cross section & 96.07 & 96.52 & 96.56 & 96.58 \\ 
Statistical error & 0.18 & 0.15 & 0.16 & 0.15 \\ 
Experimental error  & 0.85 & 0.94 & 0.88 & 0.89 \\ 
Extrapolation error  & 0.31 & 0.42 & 0.23 & 0.23 \\ \hline
Total  error & 0.92 & 0.98 & 0.93 & 0.93 \\ \hline \hline
    \end{tabular}
  \caption{The total cross section and uncertainties for four different 
   $t$-reconstruction methods. The nominal results are based on the subtraction method, quoted in the second column.}
  \label{tab:sigma_tot_syst_summary}
  \end{center}
\end{table}
The results from the nominal subtraction method are in good agreement with the other methods, considering the 
uncorrelated uncertainty of 0.3--0.4 mb. The alternative methods are correlated through the common use 
of the local angle variable. 

Further stability checks are carried out in order to cross-check the fitting method. A fit using only the covariance 
matrix of statistical uncertainties yields $\sigmatot = 96.34 \; \pm \mbox{0.07} \; (\mbox{stat.})$
in good agreement with the results from the profile fit Eq.~(\ref{eq:profilefit}). 
The same fit with only statistical uncertainties was also performed for the two arms of ALFA independently 
and gave consistent results within one standard deviation of the statistical uncertainty. The data sample was split into 
ten sub-periods with roughly equal numbers of selected events and no dependence of the measured
value of $\sigmatot$ on time was observed. 
Also, the data from the three different bunches were investigated independently 
and found to give consistent results. 
Finally the stability of the analysis was tested by a wide variation of
the event selection cuts.   
The largest change of $\sigmatot$ with  
these cut variations was observed for the cut on the correlation between $x$ and $\theta_x$. That produced a change of 
$\pm 0.3$ mb, well within the $t$-dependent experimental systematic uncertainty of about $0.5$ mb. 
Several alternative parameterizations 
~\cite{WestAndYennie, Block_and_Cahn_curvature, Selyugin, KFK, PhillipsAndBarger,Fagundes, BourrelyAndSoffer} of the 
differential elastic cross section, including non-exponential forms at large $t$, were used to fit the spectrum 
up to $-t=0.3 \GeV^2$ in order to assess the impact on the value of the total cross section. The RMS of the values 
obtained is $0.28$ mb, in good agreement with the quoted extrapolation uncertainty of $0.31$ mb assigned to the simple 
exponential form. 

\begin{figure}[h!]
  \centering
   \includegraphics[width=0.48\textwidth]{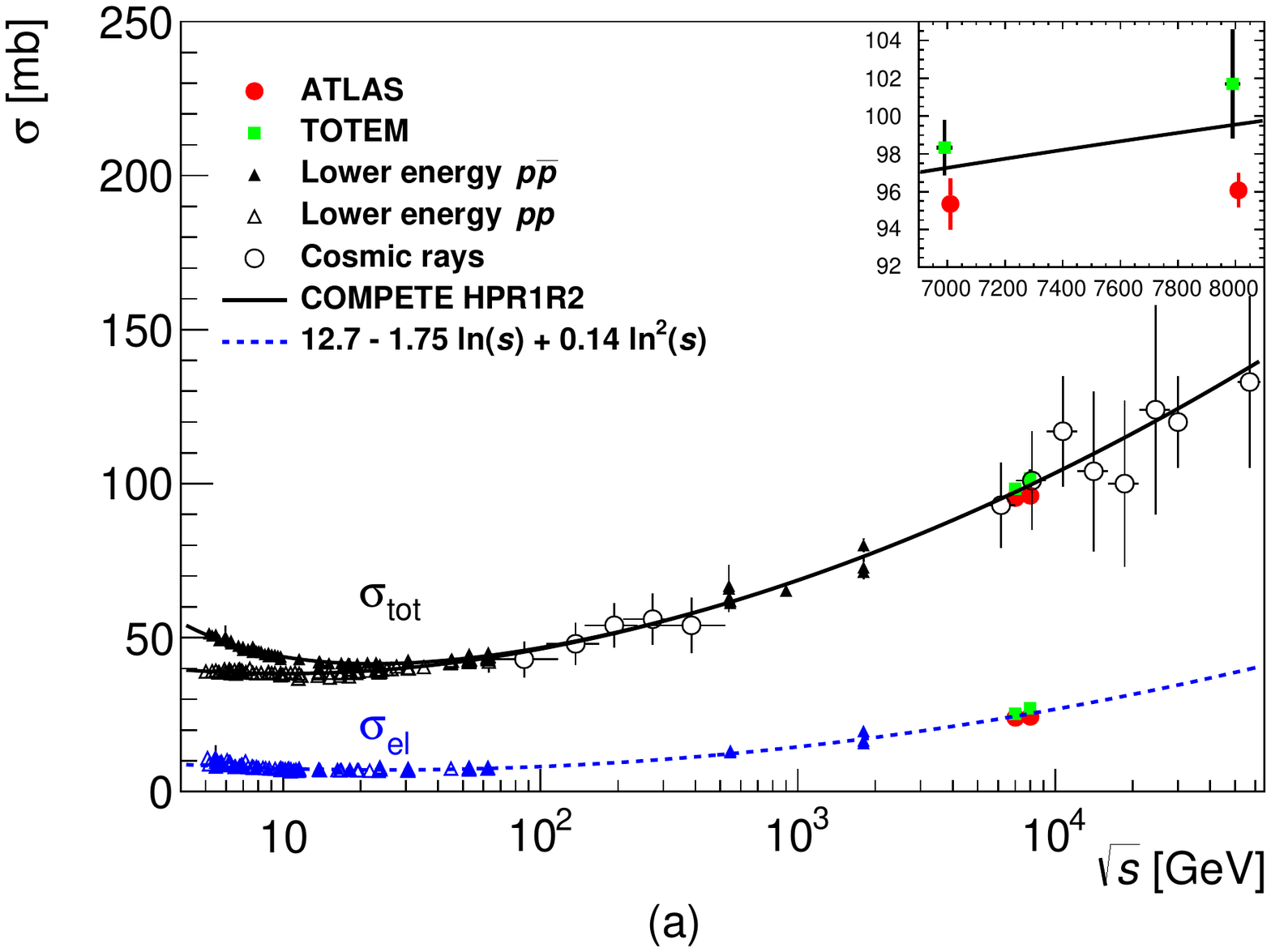} 
   \includegraphics[width=0.48\textwidth]{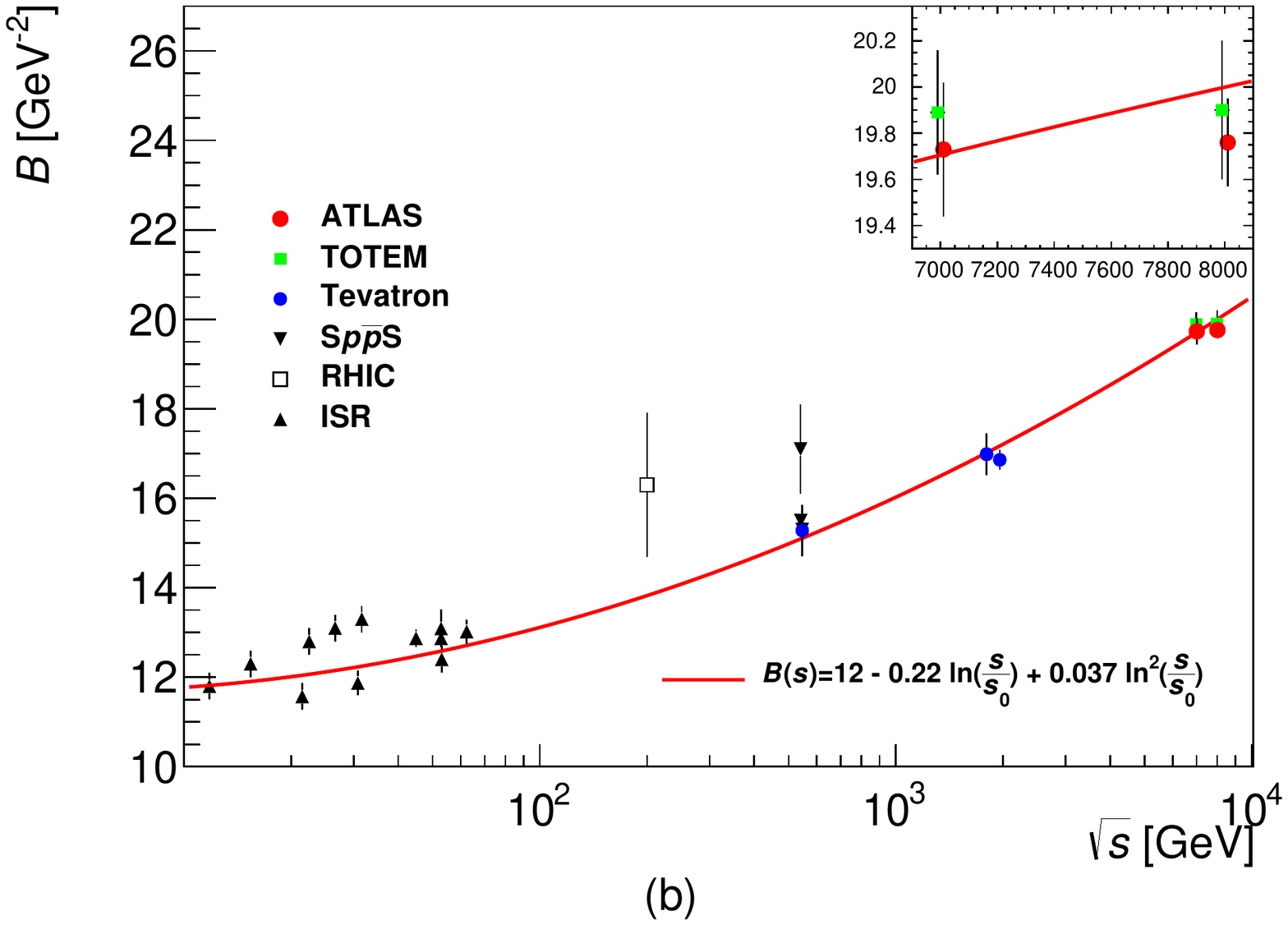}
  \caption{(a) Comparison of total and elastic cross-section measurements presented here with other
 published measurements~\cite{PDG_2014,TOTEM_second,Auger,ARGO-YBJ,AKENO,FlysEye} and model predictions as a function 
 of the centre-of-mass energy. (b) Comparison of the measurement of the nuclear slope $B$ presented here with other
 published measurements at the ISR, at the S$p\bar{p}$S, at RHIC, at the Tevatron  
and with the measurement from TOTEM at the LHC. 
The red line shows a model calculation \cite{Scheg}, which contains a linear term and quadratic term in $\ln s$.} 
  \label{fig:CrossSectionS}
\end{figure}

The TOTEM Collaboration exploited data from the same LHC fill for a measurement of $\sigmatot$ using the 
luminosity-independent method. Their result is $\sigmatot = 101.7 \pm 2.9$ mb \cite{TOTEM_8TeV}, higher than 
the measurement presented here. The difference corresponds to $1.9 \sigma$ assuming uncorrelated uncertainties. 
Better agreement is observed in the nuclear slope measurement, where TOTEM reports $B=19.9 \pm 0.3 \GeV^{-2}$, 
a value very close to the present result $B=19.74 \pm 0.19 \GeV^{-2}$, which indicates that the difference is confined to 
the normalization. 
The measurements of ATLAS and TOTEM are compared to measurements at lower energy and to a global  
fit ~\cite{PDG_2014} in Figure~\ref{fig:CrossSectionS}(a) for $\sigmatot$ and in Figure~\ref{fig:CrossSectionS}(b) for $B$.
TOTEM also reported evidence of non-exponential behaviour of the differential elastic cross section 
\cite{TOTEM_nonexp} in the $-t$-range below $0.2\GeV^2$, where deviations from the single exponential form 
of the order of one percent are observed. Such effects cannot be substantiated with this data set 
because their size is below the systematic uncertainties of the present measurement.

As well as the total cross section, the total integrated elastic cross section can
be calculated,  
provided that the 
Coulomb amplitude is neglected. In this case, $\sigmael$ can be obtained from the formula 
\begin{linenomath*}
\begin{equation}
\label{eq:elastic_extrapolation}
 \sigmael \; = \; 
\frac{\sigmatot^2}{B} \; \frac{1+\rho^2}{16\pi(\hbar c)^2} \; ,
\end{equation}
\end{linenomath*}
and the result is $\sigma_{\mathrm{el}} = 24.33 \pm 0.04 \; (\mbox{stat.}) \pm 0.39 \; (\mbox{syst.}) \; \mbox{mb}$. 
The measured integrated elastic cross section in the fiducial range from $-t=0.009$ \GeV$^2$ to $-t=0.38$ \GeV$^2$
corresponds to $80\%$ of this total elastic cross section 
$\sigmael^{\mathrm{observed}} = 19.67 \pm 0.02 \;(\mbox{stat.}) \pm 0.33 \; (\mbox{syst.}) \; \mbox{mb}$.
The total inelastic cross section is determined by subtraction of the total elastic cross section 
from the total cross section. The resulting value is 
$\sigmainel = 71.73 \pm 0.15 \; (\mbox{stat.}) \pm 0.69 \; (\mbox{syst.}) \; \mbox{mb}$.

\FloatBarrier
\section{Conclusion}
\label{sec:conclusion}
ATLAS has performed a measurement of the total cross section from elastic $pp$ scattering at $\rts=8 \TeV$.   
The measurement is based on $500$ $\mu$b$^{-1}$ of collision data collected in a high-$\betastar$ run at the LHC 
in 2012 with the ALFA Roman Pot sub-detector. The optical theorem 
is used to extract the total cross section from the differential elastic cross section by extrapolating 
$t \rightarrow 0$. The differential cross section  is also used to determine the nuclear slope. 
The analysis uses data-driven methods to determine relevant 
beam optics parameters and event reconstruction efficiency, and to tune the simulation. 
The detailed evaluation of the associated systematic uncertainties is supplemented by a comparison of 
$t$-reconstruction methods with different sensitivities to beam optics. 
The absolute luminosity for this run is determined in a dedicated analysis, 
taking into account the  special conditions with a very low number of interactions per bunch crossing. 
The total cross section at $\rts=8 \TeV$ is determined to be
\begin{eqnarray*}
\sigmatot(pp\rightarrow X) & = &  \mbox{96.07} \; \pm \mbox{0.18} \; (\mbox{stat.}) \pm \mbox{0.85} \; (\mbox{exp.})  \pm \mbox{0.31} \; (\mbox{extr.})  \; \mbox{mb} \; ,
\end{eqnarray*}
where the first error is statistical, the second accounts for all experimental systematic uncertainties and the last 
is related to uncertainties in the extrapolation $t\rightarrow 0$. 
In addition, the slope of the elastic differential cross section at small $t$ is determined to be  
$B = \mbox{19.74} \; \pm \mbox{0.05} \; (\mbox{stat.}) \pm \mbox{0.23} \; (\mbox{syst.}) \;  \GeV^{-2}$. 

The total elastic cross section is extracted from the fitted parameterization as
$\sigma_{\mathrm{el}}(pp\rightarrow pp) =  \mbox{24.33} \; \pm \mbox{0.04} \; (\mbox{stat.}) \pm \mbox{0.39} \; (\mbox{syst.}) \; \mbox{mb}$
and the inelastic cross section is obtained by subtraction from the total cross section as 
$\sigma_{\mathrm{inel}} =  \mbox{71.73} \; \pm \mbox{0.15} \; (\mbox{stat.}) \pm \mbox{0.69} \; (\mbox{syst.})\;  \mbox{mb}$.  
The measurements at $8\tev$ are significantly more precise than the previous measurements at $7\tev$ because 
 of the smaller luminosity uncertainty and a larger data sample.

\section*{Acknowledgements}
\input{Acknowledgements}

\printbibliography
\newpage
\input{atlas_authlist}

\end{document}

%% file: STDM-2015-22-metadata.tex
\AtlasTitle{Measurement of the total cross section from elastic 
scattering in $pp$ collisions at $\rts=8\TeV$  with the ATLAS detector}

\date{\today}

\PreprintIdNumber{CERN-EP-2016-158}

\AtlasJournal{Phys.\ Lett.\ B.}

\AtlasAbstract{%
A measurement of the total $pp$ cross section at the LHC at $\rts=8\TeV$ is presented. 
An integrated luminosity of $500$ $\mu$b$^{-1}$ was accumulated 
in a special run with high-$\betastar$ beam optics 
to measure the differential elastic cross section as a function of 
the Mandelstam momentum transfer variable $t$. The measurement is performed with the ALFA sub-detector of 
ATLAS. Using a fit to the differential elastic cross section in the $-t$ range from $0.014~\GeV^2$ to $0.1~\GeV^2$ to 
extrapolate $t\rightarrow 0$, the total cross section, $\sigmatot(pp\rightarrow X)$,  
is measured via the optical theorem to be
\begin{equation*}
\sigmatot(pp\rightarrow X) =  \mbox{96.07} \; \pm 0.18 \; ({\mbox{stat.}}) \pm 0.85 \; 
({\mbox{exp.}}) \pm 0.31 \; (\mbox{extr.})  \; \mbox{mb} \,, 
\end{equation*}
where the first error is statistical, the second accounts for all experimental systematic uncertainties and the last 
is related to uncertainties in the extrapolation $t\rightarrow 0$. 
In addition, the slope of the exponential function describing the elastic cross section at small $t$ is determined to be $B
 = 19.74 \pm 0.05 \; ({\mbox{stat.}}) \pm 0.23  \; ({\mbox{syst.}}) \; \mbox{\GeV}^{-2}$.    
  
}

%% file: Acknowledgements.tex

We thank CERN for the very successful operation of the LHC, as well as the
support staff from our institutions without whom ATLAS could not be
operated efficiently.
We are indebted to the beam optics development team, led by H.~Burkhardt, for the
design, commissioning and thorough operation of the high-$\betastar$ optics in 
dedicated LHC fills. 

We acknowledge the support of ANPCyT, Argentina; YerPhI, Armenia; ARC, Australia; BMWFW and FWF, Austria; ANAS, Azerbaijan; SSTC, Belarus; CNPq and FAPESP, Brazil; NSERC, NRC and CFI, Canada; CERN; CONICYT, Chile; CAS, MOST and NSFC, China; COLCIENCIAS, Colombia; MSMT CR, MPO CR and VSC CR, Czech Republic; DNRF and DNSRC, Denmark; IN2P3-CNRS, CEA-DSM/IRFU, France; GNSF, Georgia; BMBF, HGF, and MPG, Germany; GSRT, Greece; RGC, Hong Kong SAR, China; ISF, I-CORE and Benoziyo Center, Israel; INFN, Italy; MEXT and JSPS, Japan; CNRST, Morocco; FOM and NWO, Netherlands; RCN, Norway; MNiSW and NCN, Poland; FCT, Portugal; MNE/IFA, Romania; MES of Russia and NRC KI, Russian Federation; JINR; MESTD, Serbia; MSSR, Slovakia; ARRS and MIZ\v{S}, Slovenia; DST/NRF, South Africa; MINECO, Spain; SRC and Wallenberg Foundation, Sweden; SERI, SNSF and Cantons of Bern and Geneva, Switzerland; MOST, Taiwan; TAEK, Turkey; STFC, United Kingdom; DOE and NSF, United States of America. In addition, individual groups and members have received support from BCKDF, the Canada Council, CANARIE, CRC, Compute Canada, FQRNT, and the Ontario Innovation Trust, Canada; EPLANET, ERC, FP7, Horizon 2020 and Marie Sk{\l}odowska-Curie Actions, European Union; Investissements d'Avenir Labex and Idex, ANR, R{\'e}gion Auvergne and Fondation Partager le Savoir, France; DFG and AvH Foundation, Germany; Herakleitos, Thales and Aristeia programmes co-financed by EU-ESF and the Greek NSRF; BSF, GIF and Minerva, Israel; BRF, Norway; Generalitat de Catalunya, Generalitat Valenciana, Spain; the Royal Society and Leverhulme Trust, United Kingdom.

The crucial computing support from all WLCG partners is acknowledged gratefully, in particular from CERN, the ATLAS Tier-1 facilities at TRIUMF (Canada), NDGF (Denmark, Norway, Sweden), CC-IN2P3 (France), KIT/GridKA (Germany), INFN-CNAF (Italy), NL-T1 (Netherlands), PIC (Spain), ASGC (Taiwan), RAL (UK) and BNL (USA), the Tier-2 facilities worldwide and large non-WLCG resource providers. Major contributors of computing resources are listed in Ref.~\cite{ATL-GEN-PUB-2016-002}.

%% file: atlas_authlist.tex
\begin{flushleft}
{\Large The ATLAS Collaboration}

\bigskip

M.~Aaboud$^\textrm{\scriptsize 135d}$,
G.~Aad$^\textrm{\scriptsize 86}$,
B.~Abbott$^\textrm{\scriptsize 113}$,
J.~Abdallah$^\textrm{\scriptsize 64}$,
O.~Abdinov$^\textrm{\scriptsize 12}$,
B.~Abeloos$^\textrm{\scriptsize 117}$,
R.~Aben$^\textrm{\scriptsize 107}$,
O.S.~AbouZeid$^\textrm{\scriptsize 137}$,
N.L.~Abraham$^\textrm{\scriptsize 149}$,
H.~Abramowicz$^\textrm{\scriptsize 153}$
% ...
\bigskip
\\
% ...
$^{12}$ Institute of Physics, Azerbaijan Academy of Sciences, Baku, Azerbaijan\\
% ...
$^{64}$ University of Iowa, Iowa City IA, United States of America\\
% ...
$^{86}$ CPPM, Aix-Marseille Universit{\'e} and CNRS/IN2P3, Marseille, France\\
% ...
$^{107}$ Nikhef National Institute for Subatomic Physics and University of Amsterdam, Amsterdam, Netherlands\\
% ...
$^{113}$ Homer L. Dodge Department of Physics and Astronomy, University of Oklahoma, Norman OK, United States of America\\
% ...
$^{117}$ LAL, Univ. Paris-Sud, CNRS/IN2P3, Universit{\'e} Paris-Saclay, Orsay, France\\
% ...
$^{135}$ $^{(a)}$ Facult{\'e} des Sciences Ain Chock, R{\'e}seau Universitaire de Physique des Hautes Energies - Universit{\'e} Hassan II, Casablanca; $^{(b)}$ Centre National de l'Energie des Sciences Techniques Nucleaires, Rabat; $^{(c)}$ Facult{\'e} des Sciences Semlalia, Universit{\'e} Cadi Ayyad, LPHEA-Marrakech; $^{(d)}$ Facult{\'e} des Sciences, Universit{\'e} Mohamed Premier and LPTPM, Oujda; $^{(e)}$ Facult{\'e} des sciences, Universit{\'e} Mohammed V, Rabat, Morocco\\
% ...
$^{137}$ Santa Cruz Institute for Particle Physics, University of California Santa Cruz, Santa Cruz CA, United States of America\\
% ...
$^{149}$ Department of Physics and Astronomy, University of Sussex, Brighton, United Kingdom\\
% ...
$^{153}$ Raymond and Beverly Sackler School of Physics and Astronomy, Tel Aviv University, Tel Aviv, Israel\\
% ...
$^{a}$ Also at Department of Physics, King's College London, London, United Kingdom\\
$^{b}$ Also at Institute of Physics, Azerbaijan Academy of Sciences, Baku, Azerbaijan\\
$^{c}$ Also at Novosibirsk State University, Novosibirsk, Russia\\
$^{d}$ Also at TRIUMF, Vancouver BC, Canada\\
$^{e}$ Also at Department of Physics {\&} Astronomy, University of Louisville, Louisville, KY, United States of America\\
% ...
\end{flushleft}
